\documentclass{article}
\usepackage{ijcai25}

\usepackage{wrapfig}
\usepackage{float}
\usepackage{times}
\newcommand{\citet}[1]{\citeauthor{#1} \shortcite{#1}}
\usepackage{soul}
\usepackage{url}
\usepackage[hidelinks]{hyperref}
\usepackage[utf8]{inputenc}
\usepackage[small]{caption}
\usepackage{graphicx}
\usepackage{multirow}
\usepackage{amsmath}
\usepackage{amsthm}
\usepackage{booktabs}
\usepackage{algorithm}
\usepackage{tabularx}
\usepackage[noend]{algpseudocode}
\usepackage{amsmath}
\usepackage{enumitem}

\usepackage[T1]{fontenc} % add special characters (e.g., umlaute)
\usepackage[utf8]{inputenc} % set utf-8 as default input encoding
\hypersetup{
    colorlinks=true,
    linkcolor=blue,
    citecolor=blue,
    filecolor=magenta,      
    urlcolor=cyan,
    pdftitle={Overleaf Example},
    pdfpagemode=FullScreen,
    }

    \usepackage{subcaption}

\usepackage{amssymb}
\usepackage{pifont}% http://ctan.org/pkg/pifont
\usepackage{multirow}
\usepackage[symbol]{footmisc}
\usepackage{booktabs}
\usepackage[table,xcdraw]{xcolor}
\usepackage{import}

\definecolor{lightgray}{RGB}{230,230,230}
\newcommand{\grayline}{\arrayrulecolor{lightgray}\hline\arrayrulecolor{black}}
\usepackage[switch]{lineno}

\title{RenderBox: Expressive Performance Rendering with Text Control}

\author{
Huan Zhang$^1$
\and
Akira Maezawa$^2$\and
Simon Dixon$^1$\\
\affiliations
$^1$Centre for Digital Music, Queen Mary University of London\\
$^2$MINA Lab, R\&D Division, Yamaha Corporation\\
\emails
huan.zhang@qmul.ac.uk
}
\begin{document}

\maketitle

\begin{abstract}
    Expressive music performance rendering involves interpreting symbolic scores with variations in timing, dynamics, articulation, and instrument-specific techniques, resulting in performances that capture musical can emotional intent. We introduce RenderBox, a unified framework for text-and-score controlled audio performance generation across multiple instruments, applying coarse-level controls through natural language descriptions and granular-level controls using music scores. Based on a diffusion transformer architecture and cross-attention joint conditioning, we propose a curriculum-based paradigm that trains from plain synthesis to expressive performance, gradually incorporating controllable factors such as speed, mistakes, and style diversity. 
    RenderBox achieves high performance compared to baseline models across key metrics such as FAD and CLAP, and also tempo and pitch accuracy under different prompting tasks. Subjective evaluation further demonstrates that RenderBox is able to generate controllable expressive performances that sound natural and musically engaging, aligning well with prompts and intent.
\end{abstract}

\section{Introduction}

A trained musician can take a piece of music and interpret it in their own way, moulding and varying the emotional expression of the piece by subtly changing performance parameters. 
Parametric dimensions include timing, dynamics, articulation, and instrument-specific techniques such as bowing for strings, breath control for winds, or pedal usage for pianos \cite{CancinoChacon2018ComputationalModels,palmer1996anatomyexpression}.

\begin{figure}
    \centering
    \includegraphics[width=\linewidth]{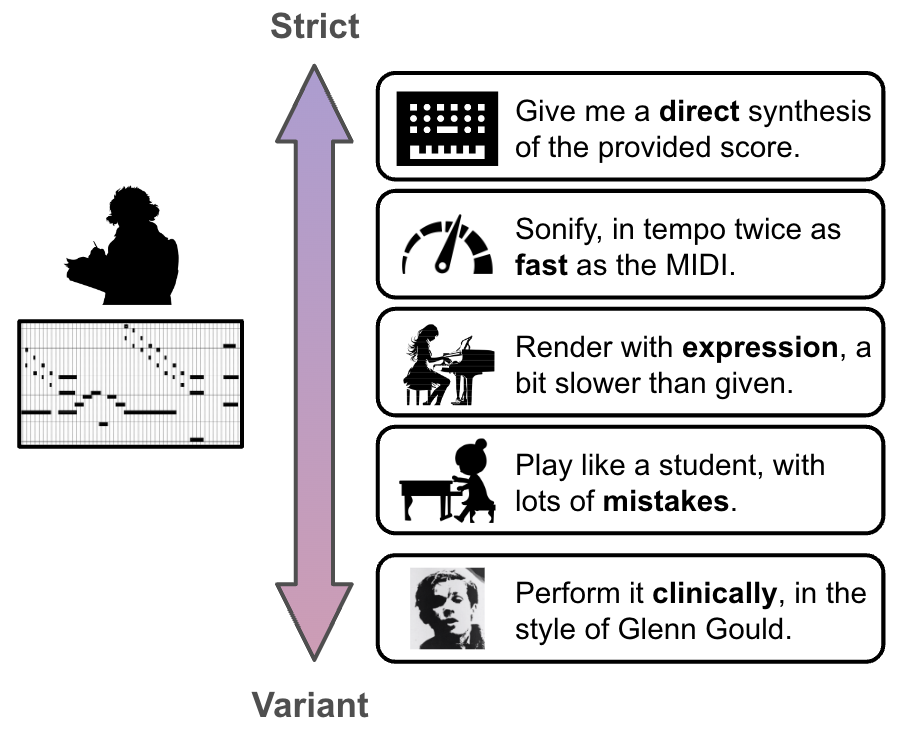}
    \caption{An overview of the performance space proposed by our paradigm, progressively from strict to variant relative to the input MIDI score.}
    \label{fig:renderbox}
\end{figure}

Studying such expression patterns has long been of keen interest to musicians, educators and researchers, and it presents a compelling inquiry into exploring whether such intricate expressions can be accurately encapsulated and replicated by computational systems \cite{Hashida2008Rencon:Systems}. Named performance rendering as a task, it has applications in technology-mediated music education and interactive music systems for entertainment and creative assistance.

Just as human speech varies in terms of accent, tone and pace, the space of performances is diverse, ranging from the mistake-prone playing of amateur players to highly personal interpretations of virtuosi. While there have been attempts to condition performance generation with controls on tempo \cite{Borovik2023ScorePerformerControl,Rhyu2022SketchingLearning} and perceptual features \cite{Zhang2024DExterModels}, enforcing flexible, multi-dimensional control via natural language has not been addressed in the performance rendering task. Here, we present \textbf{RenderBox}, a unified performance rendering framework that bridges symbolic music scores and natural language descriptions to generate expressive and controllable audio performances across multiple instruments. Our contributions\footnote{Demo page: \url{https://renderbox-page.vercel.app/}} are as follows: 
\begin{enumerate}
    \item We propose the first text-and-score controlled audio performance generation model that supports multiple instruments, enforcing coarse (text-based expressive direction) and granular (MIDI (Musical Instruments Digital Interface) score) controls into the expressive performance realm.  
    \item We present a training method that enables the model to generate performance audio signals of varying "performance variability," such as playing styles, speed, ornamentation and inclusion of mistakes.  This is achieved by applying curriculum learning with carefully designed stages of performance variability, by feeding data with progressive difficulty during training.
    \item Besides demonstrating superior performance in both objective and subjective evaluations, our model learnt interpretable similarity space regarding to the performance and composition styles in classical piano literature. 
\end{enumerate}

\section{Related Work}

\subsection{Controllability in Music Generation}

To align better with real-world music-making applications, enhancing the controllability of music generation has been a central topic since the introduction of large generative models. Recent music generation models \cite{Copet2023SimpleGeneration,Agostinelli2023MusicLMText,Melechovsky2024MustangoGeneration} are prominently based on text controls, which enables descriptions such as genre, instrumentation and mood. As discussed by \citet{Lin2024ContentModelling}, the coarse and high-level text controls are intrinsically limited, and music creators would need them to be combined with lower-level content-based controls such as melody, chord and rhythm. 
A similar idea is used in AudioBox \cite{Vyas2023AudioboxPrompts}, in which style descriptions (pace, voice type) and text transcripts jointly condition the audio outputs. 

% conditioning mechanisms
Various conditioning mechanisms are used to control music generation. Large generative models like MusicGen \cite{Copet2023SimpleGeneration}, MusicLM \cite{Agostinelli2023MusicLMText} and Riffusion\footnote{https://www.riffusion.com/} are end-to-end systems that are trained with conditioning signals as input. JASCO \cite{Tal2024JointGeneration} features multiple conditions of text, chords, melody and drums, each passing through their own representation projections that are aligned in time. Diff-A-Riff \cite{Nistal2024DiffModels} is a latent diffusion model based on consistency autoencoder input as well as CLAP \cite{Wu2023LargeAugmentation} conditioning that is inserted via FiLM \cite{Perez2018FiLMlayer}.

Motivated by transferability and resource considerations, some approaches focus on trainable control modules that augment large, pre-trained models. Coco-Mulla \cite{Lin2024ContentModelling} employs parameter efficient fine-tuning (PEFT) by inserting adaptor joint embeddings into the self-attention layers of the MusicGen decoder. Similar module-based control is used for music editing \cite{Zhang2024Instruct-Musicgen}, building on a pre-trained music generation model. Music ControlNet \cite{Wu2023MusicGeneration} adopts the ControlNet \cite{Zhang2023AddingModels} mechanism, inserting control injection layers into a pre-trained U-Net spectrogram generator. 

Extracting their controlled information from chromagrams or piano roll matrices, models like MusicGen-Melody, Coco-Mulla and Music ControlNet are conditioned with monophonic melody and generating variations (or accompaniments) that are time-aligned with the melody. However, our pursuit of expressive performance (which focuses on Western Classical and Jazz) involves much more complex, polyphonic compositions and requires outputs that are faithful to the score notes but not necessarily time-aligned. Thus, we approach the score conditioning with MIDI tokens directly, retaining the details such as exact note onset and track instrumentation, similar to a synthesis approach \cite{Hawthorne2022Multi-instrumentDiffusion}.

\subsection{Performance Rendering}

Traditional performance rendering is the task of generating a piece of natural, human-like performance given an input score, and it has been mostly applied in the symbolic domain \cite{Bresin2002DirectorSystem,Maezawa2019RenderingRNN,Jeong2019VirtuosoNetPerformance,Zhang2024DExterModels}, with a few attempts in the audio domain \cite{Dong2022DeepSynthesis,Renault2022DifferentiableSynthesis}. 
Audio performance rendering bears much resemblance to the more popular text-to-speech (TTS) task, with the goal of faithfully reproducing a piece of transcript (text or symbolic music events) in the audio realm, while retaining acoustic or style cues such as voice timbre, pace or tempo so that the audio sounds natural. Compared to text, music is much more challenging in terms of time alignment, due to its polyphonic nature, as techniques such as the phoneme duration model \cite{Le2023VoiceboxScale} that allows fine-grained alignment control in speech synthesis are harder to transfer into a music setting. 

% recognition of the variety 
Due to the complexity of the task, the variance in the music performance space has not been explored much. A piece can be played in numerous ways, ranging from masterful interpretations by virtuosi to student performances that reflect varying levels of expertise. \citeauthor{Zhang2024LLaQoAssessment} \shortcite{Zhang2024LLaQoAssessment,Zhang2024FromPiano} highlight the importance of performance understanding and the scarcity of annotated performance data.  Our current work situates the performance rendering task within this diverse interpretative space, demonstrating effective control through text-based inputs.

\begin{table*}[ht]
    \small
    \centering
    \renewcommand{\arraystretch}{1.2} % Adjust row spacing
    \setlength{\tabcolsep}{4pt} % Adjust column spacing
    \resizebox{\linewidth}{!}{%
    \begin{tabularx}{\linewidth}{p{2.5cm} p{2cm} p{2cm} c c p{2.5cm} >{\raggedright\arraybackslash}X} 
        \toprule
        \textit{Datasets by stage} & Input MIDI & Target Audio & Duration & Instrumentation & Controls / Prompts & Repertoire / Notes \\ [0.5ex]
        \midrule
        \rowcolor[gray]{.9}
        \multicolumn{7}{l}{\textit{Stage 0: Synthesis}} \\ 
        \grayline
        (n)ASAP  
        & \parbox[t]{\linewidth}{Score \& \\ Performance} 
        & \parbox[t]{\linewidth}{Synth. Score \& \\ Perf. Audio}
        & 46h
        & Piano
        & -
        & Common Practice Period solo piano works \\
        MusicNet   
        & \parbox[t]{\linewidth}{Score \& \\ Performance} 
        & \parbox[t]{\linewidth}{Synth. Score \& \\ Perf. Audio}
        & 34h
        & Orchestral
        & -
        & Classical music with a large portion of chamber works \\
        GAPS   
        & \parbox[t]{\linewidth}{Score} 
        & \parbox[t]{\linewidth}{Synth. Score }
        & 10.5h
        & Guitar
        & -
        & Noted classical guitar works \\
        % \addlinespace
        \midrule 
        \rowcolor[gray]{.9}
        \multicolumn{7}{l}{\textit{Stage 1: Speed-controlled synthesis}} \\ 
        \grayline     
        (n)ASAP-aug   
        & Score
        & Synth. Score
        & 68.3h 
        & Piano
        & Tempo 
        & Time-varied score synthesis \\
        GAPS-aug  
        & Score
        & Synth. Score
        & 25h 
        & Guitar
        & Tempo 
        & Time-varied score synthesis \\
        MusicNet-aug  
        & Score 
        & Synth. Score
        & 51h 
        & Orchestral
        & Tempo 
        & Time-varied score synthesis \\
        %\addlinespace
        \midrule 
        \rowcolor[gray]{.9}
        \multicolumn{7}{l}{\textit{Stage 2: Expressive Performance}} \\ 
        \grayline        
        (n)ASAP  
        & Score
        & Perf. Audio
        & 48.1h
        & Piano
        & Expressive 
        & Aligned perf. segments \\
        GAPS  
        & Score
        & Perf. Audio
        & 5.3h
        & Guitar
        & Expressive 
        & Aligned perf. segments \\
        In-house-Sax  
        & Score
        & Perf. Audio
        & 2.13h
        & Saxophone
        & Expressive or not %n-expressive playing
        & Jazz standards (2 versions) \\ %with two versions of playing \\
        MusicNet   
        & Score
        & Perf. Audio
        & 30.4h
        & Orchestral
        & Expressive 
        & Aligned perf. segments \\
        BachViolin  
        & Score
        & Perf. Audio
        & 4.4h 
        & Violin
        & Expressive
        & Bach violin sonata  \\
        \midrule
        \rowcolor[gray]{.9}
        \multicolumn{7}{l}{\textit{Stage 3: Mistake-corrupted performance}} \\ 
        \grayline        
        (n)ASAP-synmis  
        & Score
        & Synth. Perf.
        & 36h 
        & Piano
        & Mistakes (no annotation)
        & Synthesized from perf.\ MIDI which is corrupted with artificial mistakes \\
        \midrule
        \rowcolor[gray]{.9}
        \multicolumn{7}{l}{\textit{Stage 4: Style-directioned Performance}} \\ 
        \grayline
        ATEPP-subset  
        & Score
        & Perf. Audio
        & 273h
        & Piano
        & Performer labels 
        & Standard piano repertoire played by 40+ pianists \\
        Con Espressione  
        & Score
        & Perf. Audio
        & 0.5h
        & Piano
        & Human-annotated expression labels 
        & 45 perf.\ of 9 excerpts %, with perceptual 
        annot.\ by 179 participants  \\
        Vienna 4x22  
        & Score
        & Perf. Audio
        & 1.2h 
        & Piano
        & Performer labels 
        & 4 short pieces played by 22 pianists each \\
        \bottomrule
    \end{tabularx}%
    }
    \caption{Aligned datasets of scores and performances, organized by curriculum level. The dataset duration is computed with regards to the total length of input MIDI segments, and also subject to the availability and validity of accurate alignment.}
    \label{tab:datasets_overview}
\end{table*}

\subsection{Curriculum Learning and Continual Learning}

In our formulation of the performance variation space, a spectrum of sub-tasks with varying difficulty naturally emerges as shown in Figure~\ref{fig:renderbox}. It ranges from the ``strict / easy'' synthesis task that does not involve any alteration of the pitch-time relationship from MIDI conditioning, to the ``deviated / hard'' performance tasks that are subject to variation in tempo and timing, and even addition and removal of pitches.  By bootstrapping the tasks on a gradual difficulty scale, simulating the performance variation space can be posited in the framework of continual learning with a curriculum structure. 

Curriculum learning \cite{Wang2021SurveyLearning} enables machine learning model training to progress gradually from easy examples to harder ones, and continual learning (CL) 
% \cite{Wang2024ComprehensiveApplication} 
enables models to learn new knowledge and preserve knowledge acquired previously from a data stream with a continuously changing distribution.  
While CL is frequently applied to classification tasks \cite{Wang2019ContinualReplay,Bhatt2024CharacterizingAnalysis,Faber2023MNISTLearning}, the application of CL in generative models 
% \cite{Zhou2024Non-compositionalLearning}
is less prominent. Suffering from potential challenges such as \textit{catastrophic forgetting}, strategies such as rehearsals, replay and regularization are employed, but often at the cost of training additional discriminators. In our case, we take a data-incremental curriculum, and we are also the first to apply CL in incremental development of control for music generation.

\section{Methodology}

\subsection{Base Architecture}

We base our conditioning model on the text-to-audio architecture from Stable-audio-open \cite{Evans2024StableOpen}: an autoencoder with a latent rate of 21.5Hz to preprocess the raw waveform sampled at 44.1kHz, a T5-based text embedding for text conditioning, and a diffusion transformer (DiT) that operates in the latent space of the autoencoder with latent sequences of length 1024 (47 seconds). 

During training, we initialize the autoencoder, text embedding module and diffusion transformer with the pre-trained weights of the original text-to-audio model. 
% diffusion
% The DiT is trained to predict a noise increment from noised ground-truth latents, [fill]
The DiT is trained to predict a noise increment from noised ground-truth latents, following the standard formulation of the v-objective. Given a latent $\mathbf{z}_t$ at time $t$, perturbed with noise $\epsilon_t$ according to a variance schedule, the model predicts the velocity $\mathbf{v}_t = \alpha_t \epsilon_t - \sigma_t \mathbf{z}_t$, where $\alpha_t$ and $\sigma_t$ are determined by the variance-preserving diffusion process. During inference, we use the DPM-Solver++ for 100 steps with classifier-free guidance (scale of 7.0).

\subsubsection{MIDI Conditioning}

The text and duration (a float input with positional embedding that is used to control output length) conditioning signals are incorporated into the base architecture via cross-attention. We also need to incorporate MIDI score input to instruct the rendering.
The input MIDI data is first converted to a MIDI-Like tokenization sequence via \texttt{note-seq}\footnote{https://github.com/magenta/note-seq/}. We use the same note event vocabulary as MT3 \cite{Hawthorne2022Multi-instrumentDiffusion}, which is based on the MIDI protocol. Specifically, there are events for Instrument (128 values), Note (128 values), On/Off (2 values), Time (512 values), End Tie Section (1 value), and EOS (1 value). The MIDI conditioning module consists of an embedding layer that projects token vocabularies into a 768-dimensional space, to which a sinusoidal positional encoding is added. Our input MIDI window is 10 seconds, which could include 300--2000 tokens depending on the note density of the piece. 

\begin{figure}
    \centering
    \includegraphics[trim={2cm 2cm 2cm 2cm}, width=\linewidth]{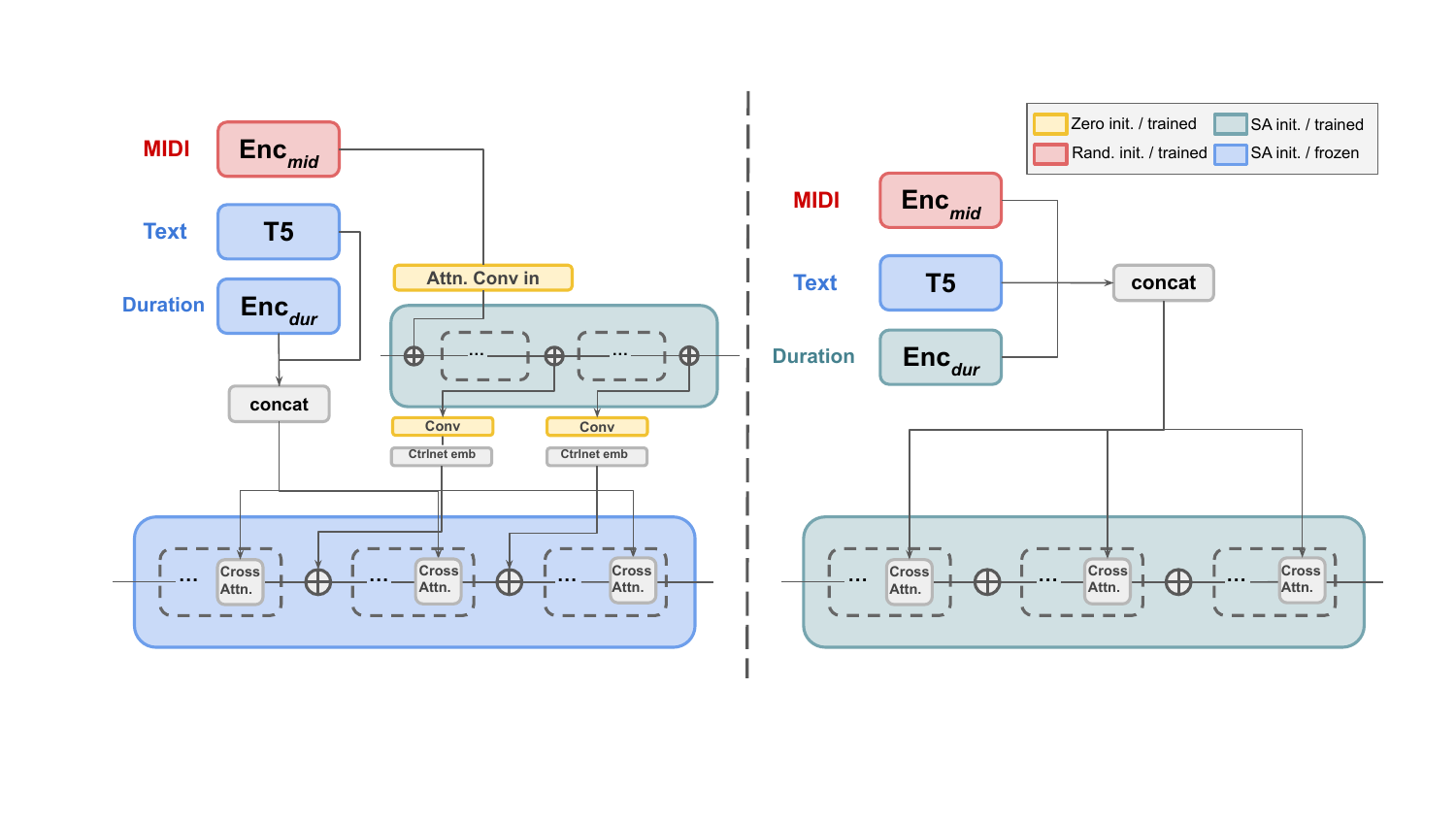}
    \caption{ControlNet conditioning (left) and concatenative cross-attention conditioning (right), with color highlighting the initialization of modules and their optimization in our experiments.}
    \label{fig:conditioning}
\end{figure}

To subsequently insert the MIDI embeddings into the DiT, we experimented with two methods:
\begin{itemize}
    \item Exp.1:
%\textbf{1)}. 
Concatenative cross-attention conditioning. The MIDI embedding is concatenated with the T5 text embedding and the duration embedding along the sequence dimension. Besides the new MIDI embedding module, all other modules and layers are initialized with the stable-audio-open weights and receive full fine-tuning (except for T5 which is frozen).
% (exp.1)
    \item Exp.2:
%\textbf{2)}. 
ControlNet conditioning based on the implementation of stable-audio-controlnet\footnote{https://github.com/EmilianPostolache/stable-audio-controlnet} \cite{Ciranni2025COCOLARepresentations}. With the main DiT and the text and duration embeddings frozen with pretrained weights, the MIDI embeddings go through the ControlNet branch which is a reduced version of DiT with around 20\% of the depth. Only the ControlNet branch and MIDI embedder are optimized. % (exp.2)
See Fig.~\ref{fig:conditioning} for an illustration of the conditioning. 
\end{itemize}

\subsection{Curriculum Training Scheduler}
\label{subsec:curriculum}

Curriculum learning sequences the learning process in a curriculum of increasing complexity tasks, which allows learning on large data collections that otherwise would be impossible to learn from scratch. As shown in Table~\ref{tab:datasets_overview}, we arrange the datasets and training targets by stages of difficulty, distributing to five stages: 
\begin{itemize}[noitemsep]
    \item Stage 0 - Synthesis, which enforces the same text prompt (20k steps) 'Synthesis'. This stage forces the model to direct its attention away from the text information to the MIDI tokens, focusing entirely on mapping MIDI events precisely to audio events.
    \item Stage 1 - Synthesis with speed change (10k steps). As human performers naturally vary tempo, this stage is an important bridge to train the model to map a stretched / squeezed time series of MIDI events in response to speed prompts (e.g., \textit{twice as fast}).
    \item Stage 2 - Expressive performance (15k steps). This stage introduces performance recordings as training target (in contrast to mechanically synthesized audio in previous stages), which involves not only global tempo change (indicated by the text prompt) but also local timing variations (such as rubato).
    \item Stage 3 - Mistake-corrupted performance (4k steps). The target is performance MIDI corrupted with artificial mistakes, such that the model learn possible pitch or rhythm deviation from the score (compared to the deviation-free performance in the previous stage) under the performance category of less-experienced player.
    \item Stage 4 - Style-directed performance (10k steps). This stage also trains with performance data, but augments the text prompt with available style directions, including performer names or expressivity annotations (i.e. \textit{calm, passionate}). 
\end{itemize}

For ablation, we also trained a version of MIDI concatenative cross-attention conditioning without the curriculum learning by mixing all stages of data (Exp.3). 

\subsection{Experiments}

All of our experiments are performed on four 80GB A100 GPUs, using 16-bit precision training. For training Exp.1 and Exp.3, we use the AdamW optimizer, with a base learning rate of 1e\text{-}5 and a scheduler including exponential ramp-up and decay. With a batch size of 108, Exp.1 trains for a total of 59k steps as stages scheduled in the previous section, and Exp.3 (no stages) is trained for 60k steps. Other training techniques such as EMA are applied, following \citet{Evans2024StableOpen}. 
For training Exp.2 of the ControlNet, we have trained 30k steps with a batch size of 108, and a learning rate of 1e\text{-}4. 

\section{Datasets}

% maybe also draw figures for data pipeline

To train our model, we have aggregated a large number of publicly available performance datasets with aligned score and audio performances, across a range of instruments. That includes (n)ASAP \cite{Peter2023AutomaticDataset}, MusicNet \cite{maman2022UnalignedWild}, GAPS \cite{Riley2024GAPSModel}, BachViolin \cite{Dong2022DeepSynthesis}, ATEPP \cite{Zhang2022ATEPPPerformance}, \textit{Con Espressione} \cite{Cancino-Chacon2020OnGame}, and Vienna 4x22\footnote{https://github.com/CPJKU/vienna4x22}. Their size, repertoire, annotations for prompts, augmentation and scheduling are listed in Table~\ref{tab:datasets_overview}.
% alignment?
In the expressive performance training of stage 2, we also utilized an in-house recorded saxophone dataset (copyrighted) of funk and swing Jazz standards from \textit{The Real Book}.

\subsection{Augmentations}
\label{subsec:augmentations}

\textbf{Speed augmentation}: For stage 1 training, we apply speed augmentation to induce the model's ability to synthesize the MIDI score timing into proportionally faster or slower audio, before learning the more flexibly varied performance timing. As shown in Table~\ref{tab:speed_tiers}, each score in the (n)ASAP, MusicNet and GAPS datasets is augmented with 5 tiers of speed ranges, in which the speed ratio is randomly sampled within the tier's range, and the text prompt is augmented with a corresponding speed keyword of common terminology. The speed-controlled re-synthesized score audio is served as the training target. We choose to not speed-augment the performances, since the artist's interpretation of timing and phrasing would be influenced by tempo \cite{Repp1996PedalInvestigation,Repp1995QuantitativeEvidence}. While the speed ranges include significant tempo shifts, they intend to help the model generalize across different expressive timing variations, rather than imply that all music would be performed at such extreme tempos.

\noindent \textbf{Mistake augmentation}: For the stage 4 training with mistake instruction, we followed the mistake taxonomy proposed by \citet{Morsi2024SimulatingLearning} and implemented several types of mistakes to augment the ASAP dataset: mistouch, asynchrony (delay or anticipation), substitution, ghost note and re-orientation (time block removal). The procedure of applying the mistakes to each segment is detailed in Algorithm~\ref{alg:mistakes}. Note that our mistake corruption does not involve adding shifts or silence on the time axis, since that would change the score-audio segment alignments.

\begin{algorithm}[ht]
\caption{Mistake Augmentation on Piece Level}
\label{alg:mistakes}
\begin{algorithmic}[1]

\State \textbf{Input:} A sequence of notes $\mathcal{N}$ with total duration $\mathcal{T}$ seconds, each with properties: pitch $\pi_n$, velocity $v_n$, start time $s_n$, and end time $e_n$.

\State \textbf{Definitions:}
\State $P_{\text{uniform}}(a, b)$: A random value uniformly sampled from the interval $[a, b]$
\State $\mathcal{U}(0, 1)$: A uniform distribution over the interval $[0, 1]$

\For{each note $n \in \mathcal{N}$}
    \State \textbf{Mistouch:}
    \If{$\mathcal{U}(0, 1) < 0.05$}
        \State Generate a new note $n'$:
        \State \hspace{1em} $\pi_{n'} \gets \pi_n + \text{choice}(\{-1, 1\})$
        \State \hspace{1em} $v_{n'} \gets 0.8 \cdot v_n$
        \State \hspace{1em} $s_{n'} \gets s_n + P_{\text{uniform}}(0.02, 0.1)$
        \State \hspace{1em} $e_{n'} \gets s_{n'} + P_{\text{uniform}}(0.1, 0.3)$
        \State Add $n'$ to $\mathcal{N}$
    \EndIf
    
    \State \textbf{Asynchrony:}
    \If{$\mathcal{U}(0, 1) < 0.2$}
        \State Shift $s_n$ and $e_n$ by $P_{\text{uniform}}(-0.7, 0.7)$
        \State Ensure $s_n \geq 0$ and $e_n \geq s_n$
    \EndIf

    \State \textbf{Substitution:}
    \If{$\mathcal{U}(0, 1) < 0.05$}
        \State $\pi_n \gets \pi_n + \text{choice}(\{-1, 1\})$
    \EndIf

    \State \textbf{Ghost Notes:}
    \If{$\mathcal{U}(0, 1) < 0.05$}
        \State Remove $n$ from $\mathcal{N}$
    \EndIf
\EndFor

\State \textbf{Time Block Removal:}
\For{$k = 0$ to $\lfloor \frac{\mathcal{T}}{5} \rfloor$}
    \State $t_{\text{start}} \gets 5k + P_{\text{uniform}}(0, 5)$
    \State $t_{\text{end}} \gets t_{\text{start}} + P_{\text{uniform}}(0.2, 0.5)$
    \State Remove all $n \in \mathcal{N}$ where $t_{\text{start}} \leq s_n < t_{\text{end}}$
\EndFor

\end{algorithmic}
\end{algorithm}

\subsection{Prompt Preparation}

Our prompts, aligning with the tasks to provide multiple stages of instructions, may include the following fields of information: \textbf{sonification type} - `\textit{synthesis}' or `\textit{performance}' (all stages), \textbf{speed keyword} (stage 1 and after), \textbf{title}, \textbf{composer}, \textbf{instrumentation}, \textbf{mistake} (stage 3), \textbf{performer ID} (stage 4), \textbf{expression label} (stage 4), subject to the availability of this information in the metadata as shown in Table~\ref{tab:datasets_overview}. 

The speed keywords (Table~\ref{tab:speed_tiers}) are first utilized in stage 1 with the speed-augmented synthesis as described in section~\ref{subsec:augmentations}. For the later stages of performance data, we incorporate a tempo prompt by estimating the length ratio of the aligned performance window to the reference score. For example, if the aligned performance window is 17 seconds of a given 10-second MIDI score, we simplify the tempo ratio as 1.7 and supply a prompt keyword of \textit{Considerably slower}. For our fixed-length (10s) input window, the performance ratios in our datasets are roughly between 0.4 and 2.2.
For other available metadata such as title, composer and instrumentation, we also choose to optionally include them (with random dropout of 0.5) as these labels would facilitate rendering by giving extra context about the MIDI piece. 

The prompt field values are specified in a comma-separated list in any order (e.g. ``\textit{a bit slower, expressive performance, Bach, Piano}'' in stage 2, or ``\textit{expressive performance, style of Vladimir Ashkenazy, Etude Op.25 No.11, Chopin, notably faster}'' in stage 4).

% Freesound examples include natural language descriptions as well as the title of the recording and tags. FMA music examples include metadata like year, genres, album, title, and artist. 

% We generate text prompts from the metadata by concatenating a random subset of the metadata as a string.

% This allows for specific properties to be specified during inference,
% while not requiring these properties to be present at all times. For
% metadata-types with a list of values, like tags or genres, we shuffle the list. As a result, we perform a variety of random transformations
% to the resulting string, shuffling the order and also transforming between upper and lower case. For half of the FMA prompts, we include the metadata-type (e.g., artist or album) and join them with
% a comma (e.g., “year: 2021, artist: dadabots, album: can’t play
% instruments, title: pizza hangover”). For the other half, we do not
% include the metadata-type and join them with a comma (e.g., “dadabots, can’t play instruments, pizza hangover, 2021”).

\begin{table}
\small
\centering
\renewcommand{\arraystretch}{1.2} % Adjust row height
\begin{tabular}{@{}l c p{4.5cm}@{}}
\toprule
\textbf{Tier} & \textbf{Dur-Range} & \textbf{Speed keyword for prompt} \\ 
\midrule
Very Slow & (1.8, 2.2) & \textit{Twice as slow as} \newline \textit{Significantly slower} \newline \textit{About half the speed o}f \\ 
\addlinespace
{Slow} & (1.5, 1.8) & \textit{Considerably slower} \newline \textit{Moving slower} \\ 
\addlinespace
Slightly Slow & (1.2, 1.5) & \textit{A bit slower than score} \newline \textit{Just under the score’s pace} \newline \textit{Slightly behind the intended pace} \\ 
\addlinespace
Neutral & (0.8, 1.2) & \textit{At the original speed }\newline \textit{In line with the score’s tempo} \\ 
\addlinespace
Slightly Fast & (0.6, 0.8) & \textit{A bit faster} \newline \textit{Just above the score’s speed} \newline \textit{Slightly faster than score} \\ 
\addlinespace
Fast & (0.4, 0.6) & \textit{Notably faster} \newline \textit{Well beyond the original tempo} \\ 
\bottomrule
\end{tabular}
\caption{Speed prompt keywords arranged in tiers.}
\label{tab:speed_tiers}
\end{table}

\begin{table*}[t]
% \small
\centering
\renewcommand{\arraystretch}{1.5} % Adjust row spacing
\setlength{\tabcolsep}{3pt} % Adjust column spacing
\resizebox{\textwidth}{!}{%
\begin{tabular}{lcccccccccccc}
\toprule
\multirow{2}{*}{Model} & \multicolumn{4}{c}{\textbf{Stage 0 - Synthesis}} & \multicolumn{4}{c}{\textbf{Stage 2 - Performance}} & \multicolumn{4}{c}{\textbf{Stage 4 - Directed Performance}} \\
\cmidrule(lr){2-5} \cmidrule(lr){6-9} \cmidrule(lr){10-13}
 & $\mathrm{FAD}_{openl3}$ $\downarrow$ & $\mathrm{CLAP}_{score}$ $\uparrow$ & Pitch $\uparrow$ & Tempo $\downarrow$
 & $\mathrm{FAD}_{openl3}$ $\downarrow$ & $\mathrm{CLAP}_{score}$ $\uparrow$ & Pitch $\uparrow$ & Tempo $\downarrow$ 
 & $\mathrm{FAD}_{openl3}$ $\downarrow$ & $\mathrm{CLAP}_{score}$ $\uparrow$ & Pitch $\uparrow$ & Tempo $\downarrow$ \\ 
\midrule
MusicGen-Melody  & 1017.5 & 0.194 & 0.496 & 2.036 & 936.88 & 0.263 & 0.522 & 11.12 & 931.83 & 0.216 & 0.523 & 23.14 \\
Coco-Mulla  & 984.3 & \textbf{0.260} & 0.739 & 1.745 & 972.65 & 0.212 & 0.704 & 11.65 & 994.09 & 0.231 & 0.674 & 23.30  \\
MIDI-DDSP   & 1436.4 & 0.148 & 0.730 & \textbf{0.006} & 1434.5 & 0.264 & 0.681 & 10.80 & - & - & - & - \\
% \rowcolor[gray]{.9}
\textbf{RenderBox-Stage0} & \textbf{839.3} & 0.174 & \textbf{0.794} & 0.038 & - & - & - & - & - & - & - & - \\ 
\textbf{RenderBox-Stage2} & 934.9 & 0.179 & 0.780 & 0.013 & \textbf{562.58} & \textbf{0.265} & 0.753 & \textbf{10.39} & - & - & - & - \\ 
\textbf{RenderBox-Stage4} & 1002.1 & 0.177 & 0.789 & 0.016 & 611.18 & 0.259 & \textbf{0.762} & 11.55 & \textbf{340.77} & \textbf{0.234} & \textbf{0.748} & \textbf{19.51} \\ 
\textbf{RenderBox-no-CL} & 1369.4 & 0.167 & 0.755 & 1.054 & 735.68 & 0.215 & 0.697 & 12.04 & 504.15 & 0.202 & 0.695 & 26.38 \\
\textbf{RenderBox-ControlNet-no-CL} & 1082.7 & 0.175 & 0.655 & 1.650 & 1127.2 & 0.124 & 0.630 & 14.26 & 781.58 & 0.144 & 0.610 & 26.06 \\ 
\bottomrule
\end{tabular}%
}
\caption{Comparison of models across the three main stages, using FAD, CLAP, Chroma, and Tempo scores as metrics. Each stage features different prompting, MIDI conditioning and ground truth data as shown in Table~\ref{tab:datasets_overview}.
For MIDI-DDSP, the evaluation is only performed on the MusicNet and BachViolin subsets as the model is restricted to chamber instruments.}
\label{tab:results}
\end{table*}

\section{Evaluation}

% inference specifications

We compare our work with the following three models. \textbf{MusicGen-melody} is the melody-conditioned version of MusicGen \cite{Copet2023SimpleGeneration}. Given that the melody conditioning is implemented via an audio input, we synthesize the MIDI score\footnote{soundfont: https://github.com/mrbumpy409/GeneralUser-GS} for MusicGen-melody as score conditioning. \textbf{Coco-Mulla} is a MusicGen-based conditioning model which enforces more external controls such as drum track, chord symbols and reference MIDI. In our usage, we supply null values for the drum track and chord symbols, taking it as a MIDI-and-text conditioned generation. \textbf{MIDI-DDSP} is an expressive synthesizer framework dedicated to string and wind instruments that is able to control aspects such as brightness and vibrato. 
Evaluations are conducted in stage 0, 2, and 4 with respect to each stage's testing set and their prompting.

\begin{figure}[ht]
    \centering
    \includegraphics[width=\linewidth]{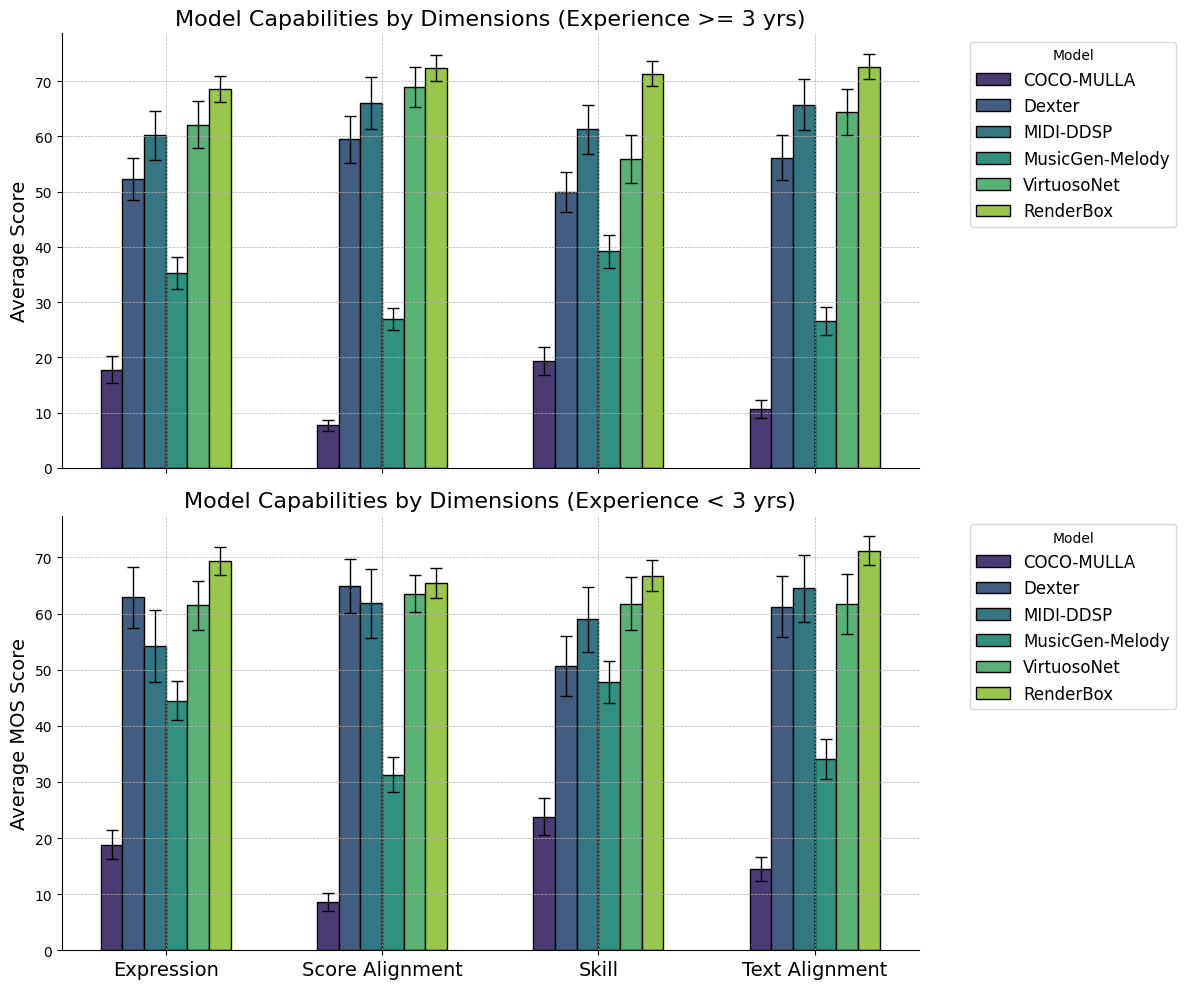}
    \caption{MOS score of the subjective evaluation on the four dimensions, separated by participant's experience.}
    \label{fig:listening_test}
\end{figure}

\subsection{Objective Evaluation}

For the audio metrics, we follow the previously established metrics \cite{Evans2024StableOpen} implemented in stable-audio-metrics, including the \textbf{Fréchet audio distance} (FAD) on \textit{OpenL3} embeddings \cite{Cramer2019LookEmbeddings} between the output and ground truth distributions, and distance in LAION-CLAP space \cite{Wu2023LargeAugmentation} between the text prompt and output.
Given the goal of performance rendering, it is crucial to enforce that the correct piece is played. To evaluate pitch-wise accuracy, we compute the \textbf{chroma similarity}, inspired by the evaluation approach used for MusicGen-Melody \cite{Copet2023SimpleGeneration}. As the output and ground truth audio are not necessarily time-aligned, we employ dynamic time warping (DTW) on the chromagrams to estimate frame correspondences for a rough audio-level alignment. Following this alignment, frame-wise \textbf{cosine similarity} is calculated between the aligned chroma features. Additionally, the DTW alignment cost is incorporated as a penalty term, scaled by a weighting factor $\lambda=10^{-3}$. 
% The overall pitch score is thus computed as:$\text{Pitch Score} = \text{Mean Cosine Similarity} - \lambda \cdot \text{Alignment Cost}$

We also measure \textbf{tempo deviation} to evaluate whether the output tempo is within the prompted tempo tier as specified in Table~\ref{tab:speed_tiers}. Given the expected tempo ratio range (for stage 0 synthesis we expect tempo not to change), the tempo deviation is computed as the percentage difference between the estimated output tempo (using \texttt{madmom}) and the score tempo adjusted by the prompt. % We use \texttt{madmom}'s tempo estimator. 

\begin{figure*}
    \centering
    \includegraphics[trim={4cm 2.2cm 3cm 7cm}, width=\linewidth]{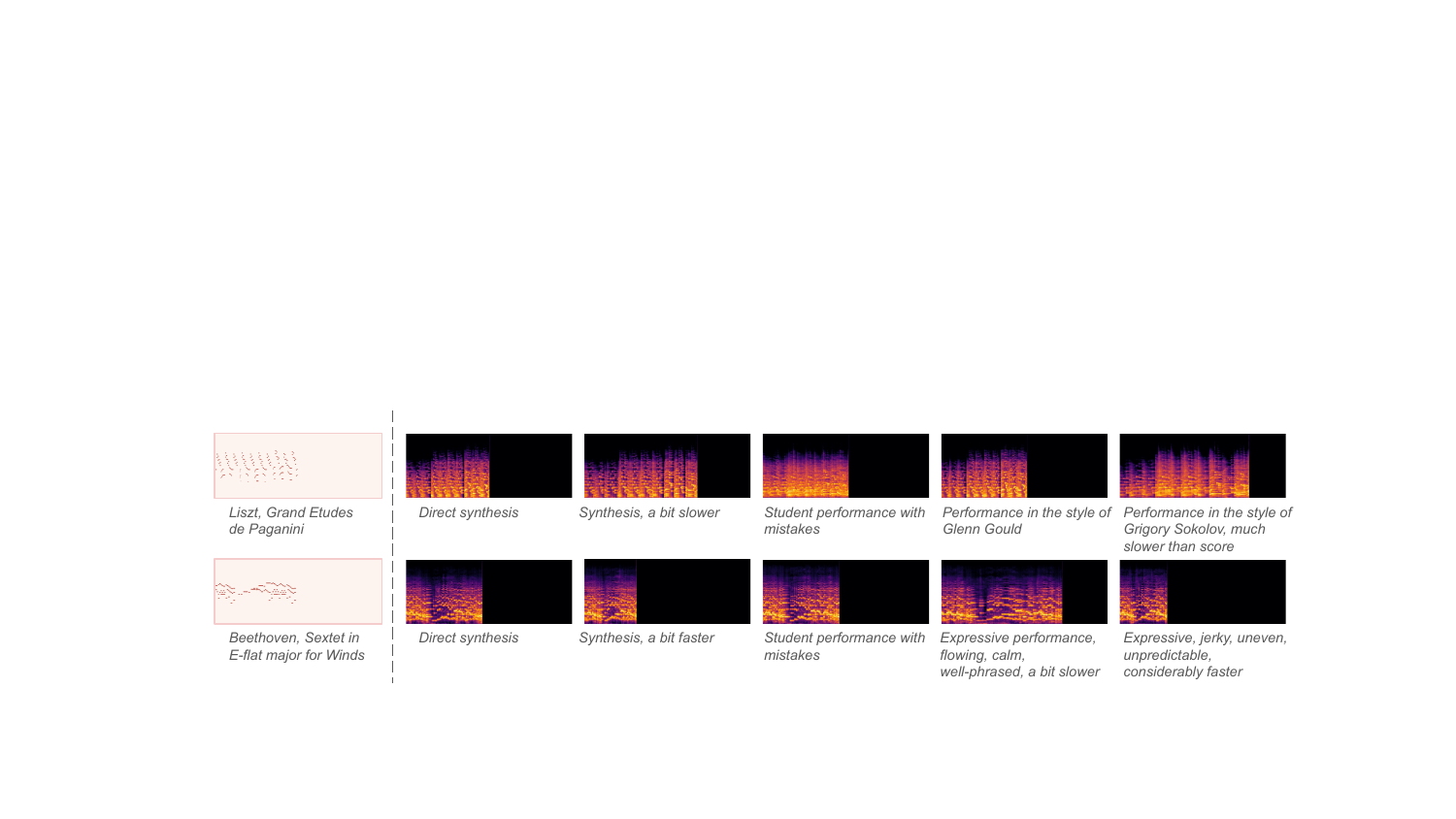}
    \caption{Input MIDI piano rolls and output spectrograms with respect to different text prompting. All visualizations are 20-second windows. }
    \label{fig:enter-label}
\end{figure*}

\subsection{Subjective Evaluation}

Besides the aforementioned baselines, we compare with two additional symbolic-output performance rendering models: VirtuosoNet \cite{Jeong2019VirtuosoNetPerformance} and DExter \cite{Zhang2024DExterModels}. The performances are synthesized from MIDI\footnote{soundfont: https://github.com/SunsetMkt/sgm\_plus\_archive}, as the focus of the subjective evaluation is on the musical content rather than audio quality.  Given the conditioning score and text prompt, test participants were asked to rate examples on a 100-point numeric scale on text alignment, music score alignment, expressivity and skill. We included nine questions, spanning six instruments as well as all the task (prompt) types in Section~\ref{subsec:curriculum}. Responses were collected from 23 participants, 10 of whom have more than 5 years' experience of instrument playing. 

The results, illustrated in Fig.~\ref{fig:listening_test}, reveal that RenderBox achieves the highest overall scores across most dimensions for both experienced and inexperienced participants. RenderBox significantly ($p<0.05$) outperforms the conditioned audio generation models Coco-Mulla and MusicGen-Melody in all four dimensions. While symbolic models (those bypassing note event prediction: DExter, VirtuosoNet, and MIDI-DDSP), achieve comparably positive feedback in score alignment, they lack the balanced performance across dimensions demonstrated by RenderBox.

In general, experienced listeners exhibit a larger difference across the models, due to their greater sensitivity to musical nuances. However, text correspondence does not seems to  significantly influence listeners' overall perception. Although RenderBox is the only model capable of speed control based on the text prompt, this feature alone does not seem to heavily impact MOS score compared to the output of the symbolic models as long as they sound musically correct. 
% when it comes to musically correct model outputs (i.e.\ the symbolic models).

\subsection{Results and Ablation}

As shown in table~\ref{tab:results}, RenderBox largely outperforms the compared baselines in most metrics, except the CLAP score and tempo deviations from the synthesis stage. [to add]
Comparison with \textbf{RenderBox-no-CL} demonstrates the effectiveness of the curriculum learning approach: Despite training with the same data with the same number of iterations, \textbf{RenderBox-no-CL} does not perform as well as the CL version from any of the stages. 
The \textit{forgetting} phenomenon is also observed in the evaluations: For the RenderBox models trained up to stage 2 and stage 4, their performance on the stage 0 synthesis task decreased dramatically, as they have been fitted with increasingly diverse distributions.

The controlnet experiment, trained with full set of data in a non-bootstrapped manner, could not outperform the main RenderBox experiments' result. The zero-convolution-inserted MIDI tokens does not yield a strong impact on the output audio compared to the text conditioning. 
% [part on controlnet - todo]

\subsection{% Observations from 
The Performer-Piece Embedding Space}

In stage 4, we directed the model to learn highly specific performance styles by fitting the model on ATEPP, a collection of virtuoso piano recordings. Although evaluating a pianist's unique style is inherently challenging due to the expertise required, the outputs of our model received very positive feedback during informal interviews with students and professors from conservatories.\footnote{We strongly encourage readers to explore our demo page.}

\begin{figure}
    \centering
    \includegraphics[width=\linewidth]{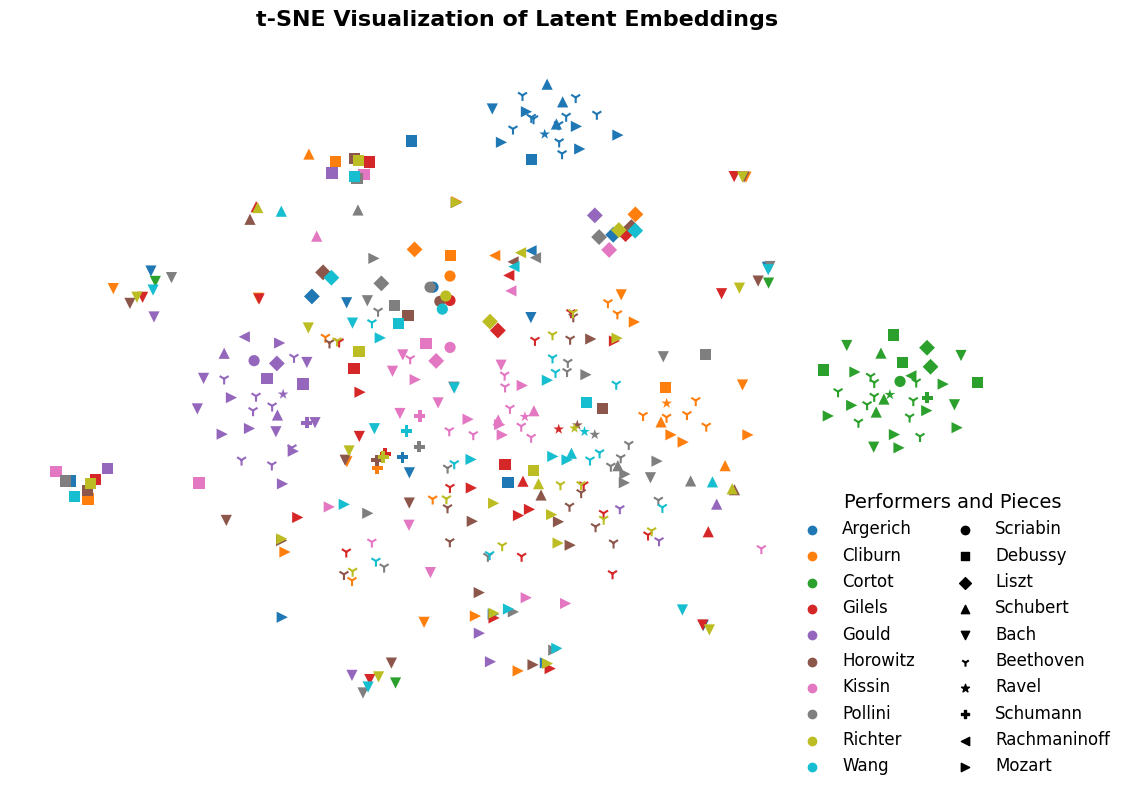}
    \caption{t-SNE visualization of generation with testing data subset, colored by performers and shaped by composers.}
    \label{fig:performer_similarity}
\end{figure}

In figure~\ref{fig:performer_similarity} we surveyed 380 test data pieces, generating with prompt ``In the style of pianist X,'' where X is one of ten famous pianists. We plot the final step's denoised latent as a t-SNE reduction (\textit{perplexity}=30).
Some pianists' styles clustered prominently regardless of the piece like \textit{Argerich}, \textit{Cortot} and \textit{Gould}. 
Composers also form clusters, indicating a greater consistency across interpretations that transcends individual performer styles, such as the \textit{Debussy} cluster, \textit{Bach} cluster and \textit{Mozart} cluster at the bottom. Within the composer clusters, we also witnessed patterns of style proximity such as \textit{Gilels} and \textit{Richter}, who are often close regardless of piece, which can be explained by the fact that they are both Russian Silver Age pianists and Neuhaus's pupils \cite{neuhaus}. Modern pianists such as \textit{Yuja Wang} feature much more spread-out interpretations according to the model.

\vspace{-0.5em}

\section{Conclusion and Future Work}

We introduced RenderBox, the first model capable of generating expressive performances with text-based control. Combining coarse language descriptions with fine-grained MIDI conditioning, RenderBox achieves flexible control for speed, mistakes, and style diversity with multiple instruments. 
% Utilizing a diffusion transformer architecture and a curriculum-based training paradigm, the model bridges symbolic scores and audio, delivering superior musical fidelity and controllable expressivity compared to existing methods.

While RenderBox demonstrates robust performance, limitations persist in its handling of acoustic quality due to the lack of detailed annotations for instrument timbres in the training data. 
Still, as the model bridges symbolic scores and audio, future creative applications can leverage annotated datasets to support instrument transfer, orchestral arrangements from symbolic scores, refinement of MIDI inputs with mistakes into polished audio outputs, and even generation with specific performance techniques.
% These directions aim to enhance RenderBox’s utility for creative and educational applications, establishing it as a versatile tool for expressive audio generation.

% limitation on acoustic side - we train on a lot of data our instruments' acoustics are not really annotated so the things like violin sound is not always consistent

% future creative applications - since we have essentially build a bridge for strong MIDI conditioning with 
% - instrument or timbre transfer or even acoustic conditioning with text prompt 
% - arrangements from like piano score to symphony recordings
% - expansion to specific performance technique , as long as there is annotation
% - mistake generation from the other side - salvage an input midi that contains mistakes and output smooth recording

% \section*{Acknowledgments}
% qm and yamaha
% The listening test is supported by the golisten platform \cite{zhang_barry_sun_hines_2021}. 

% disclaimer

% For Yamaha's side:
% This work was conducted as a part of an internship project at Yamaha Corporation.

\newpage

\bibliographystyle{named}
\bibliography{ijcai24, ref}

\end{document}